\documentclass[pra,twocolumn,superscriptaddress,showpacs,nofootinbibfloatfix,amsmath,amsfonts,amssymb]{revtex4-1}%
\usepackage{amsmath,amsfonts,amssymb,color}
\usepackage{amsthm}
\usepackage{leftidx}
\usepackage{graphicx}
\usepackage{xcolor}
\usepackage{dcolumn}
\usepackage{bm}
\usepackage{epstopdf}
\usepackage{epsfig}
\usepackage{environ}
\usepackage{pdfcomment}

\usepackage{float}
\usepackage[T1]{fontenc}
\usepackage[latin9]{inputenc}
\usepackage{setspace}
\usepackage{esint}

\begin{document}

\preprint{APS/123-QED}

\title{Dynamical characterization of non-Hermitian Floquet topological phases in one-dimension}

\author{Longwen Zhou}
\email{zhoulw13@u.nus.edu}
\affiliation{%
	Department of Physics, College of Information Science and Engineering, Ocean University of China, Qingdao, China 266100
}

\date{\today}

\begin{abstract}
Non-Hermitian topological phases in static and periodically driven
systems have attracted great attention in recent years. Finding dynamical
probes for these exotic phases would be of great importance for the
detection and application of their topological properties. In this
work, we introduce an approach to dynamically characterize
non-Hermitian Floquet topological phases in one-dimension with chiral symmetry. We show
that the topological invariants of a chiral symmetric
Floquet system can be fully determined by measuring the winding angles
of its time-averaged spin textures. We further purpose a piecewise
quenched lattice model with rich non-Hermitian Floquet topological
phases, in which our theoretical predictions are numerically demonstrated
and compared with another approach utilizing the mean chiral displacement
of a wavepacket.
\end{abstract}

\maketitle

\section{Introduction}\label{sec:Int}

Topological states of matter in non-Hermitian systems have attracted
much interest in recent years~\cite{NHTPReview1,NHTPReview2,NHTPReview3,NHTPReview4,NHTPReview5,NHTPReview6,NHTPReview7,NHTPReview8,NHTPReview9,NHTPReview10}. Theoretically, the presence of gain
and loss or nonreciprocal effects induce rich static/Floquet topological
phases~\cite{RudnerPRL2009,EsakiPRB2011,DiehlNP2011,ZhuRPA2014,MalzardPRL2015,YucePLA2015,San-Jose2016,LeePRL2016,LeykamPRL2017,JinPRA2017,XuPRL2017,KlettPRA2017,YinPRA2018,LieuPRB2018,DangelPRA2018,ZhouPRB2018,ZhouPRA2018,LiuPRL2019,LonghiPRL2019,ZhangPRA2019,HirsbrunnerPRB2019,ZhouOP2019,OkugawaPRB2019,YangPRB2019,YoshidaPRB2019,MoorsPRB2019,BudichPRB2019,CaspelSP2019} and exotic phenomena like the non-Hermitian skin effect~\cite{NHSkin1,NHSkin2,NHSkin3,NHSkin4,NHSkin5,NHSkin6,NHSkin7,NHSkin8,NHSkin9,NHSkin10,NHSkin11,NHSkin12} and unique entanglement properties~\cite{NHES1,NHES2,NHES3}, resulting
in the reformulation of topological classification schemes~\cite{Class1,Class2,Class3,Class4,Class5,Class6,Class7,Class8,Class9,Class10} and principles of bulk-edge correspondence~\cite{BBC1,BBC2,BBC3,BBC4,BBC5,BBC6,BBC7,BBC8,BBC9,BBC10,BBC11} for their description. Experimentally,
non-Hermitian topological phases have been observed in optical~\cite{Exp11,Exp12,Exp13}, photonic~\cite{Exp21,Exp22,Exp23}, topolectric circuit~\cite{Exp31}, optomechanical~\cite{Exp41,Exp42}, and mechanical systems~\cite{Exp51,Exp52}, leading to potential applications like topological energy transfer~\cite{EPTransfer1}, topological lasers~\cite{TPLaser1,TPLaser2,TPLaser3} and enhanced sensitivity in optics~\cite{EPSense1,EPSense2,EPSense3,EPSense4}.

An indispensable step in the search of non-Hermitian topological matter
is to find their defining topological signatures. One type of such
signatures can appear as topological edge states at the boundaries
of the system~\cite{Exp22,Exp52}, which may further lead to quantized or non-quantized transport coefficients. Another type of topological signature is formed by the
dynamical pattern of bulk states under external perturbations~\cite{Exp11,Exp23}. In
Ref.~\cite{ZhouarXiv2019}, it was shown that the mean chiral displacement~\cite{MCD1,MCD2,MCD3} of a wavepacket
in nonequilibrium evolution could be used to probe the
topological invariants of non-Hermitian Floquet systems in one-dimension (1d).
More recently, a dynamical classification scheme~\cite{SpinTex1} for non-Hermitian
topological matter is proposed, which allows for the extraction of topological
invariants from the time-averaged spin textures of non-Hermitian static
systems in one and two dimensions~\cite{ZhuarXiv2019}.

In this work, we provide a dynamical characterization of non-Hermitian
topological phases in 1d Floquet systems. We first
review the definition of topological winding numbers for a chiral
symmetric Floquet system and the description of its
stroboscopic dynamics in biorthogonal representations. Following that,
we propose our construction of dynamical topological invariants from
the winding angles of stroboscopic time-averaged spin textures over
the first Brillouin zone (BZ). We further show that these invariants are
equal to the topological winding numbers of the Floquet system under
generic initial conditions. Finally, we propose a piecewise
quenched 1d lattice model with rich non-Hermitian Floquet topological
phases, and verify our theory by explicit numerical simulations. Our approach therefore achieves a dynamical
characterization of non-Hermitian Floquet topological phases in 1d, with potential applications in their experimental detections.

\section{Chiral symmetric Floquet systems in 1d}\label{sec:CSFloquet}
In this section, we discuss the identification of chiral symmetry (CS) for a Floquet system and the definition of the corresponding topological invariants. The stroboscopic dynamics of a Floquet system is described by its time evolution operator over a complete driving period, i.e., $U={\cal T}e^{-\frac{i}{\hbar}\int_{0}^{T}H(t)dt}$, which is also called the Floquet operator. Here ${\cal T}$ executes the time ordering, $T$ is the driving period, and $H(t)$ is the time-dependent Hamiltonian of the system. The definition of chiral symmetry for the Floquet operator $U$ relies on the existence of a pair of symmetric time frames, which are obtained by shifting the starting time of the evolution~\cite{AsbothSTF}. To be explicit, we can first decompose a driving period $T$ into two arbitrary time durations $t_a$ and $t_b$, such that $T=t_a+t_b$. During $t_a$ and $t_b$, the evolution operators of the system can be effectively expressed as $U_a=e^{-\frac{i}{\hbar} H_a t_a}$ and $U_b=e^{-\frac{i}{\hbar} H_b t_b}$. $H_a$ and $H_b$ are the effective Hamiltonians governing the dynamics in time durations $t_a$ and $t_b$, respectively. The Floquet operator $U$ can also be written in terms of $U_a$ and $U_b$ as $U = U_b U_a$. We can then perform similarity transformations to obtain the Floquet operators in two different time frames as $U_1=U_a^{1/2}U_bU_a^{1/2}$ and $U_2=U_b^{1/2}U_aU_b^{1/2}$. Note that the three Floquet operators $\{U,U_{1},U_{2}\}$ share the same quasienergy spectrum since they are related with one another by similarity transformations. Then the Floquet system described by $U$ has CS if there exists a unitary transformation $\Gamma$, such that $\Gamma^{2}=1$ and $\Gamma U_{\alpha}\Gamma=U_{\alpha}^{-1}$ for $\alpha=1,2$. Note that the chiral symmetry defined here is the same as the chiral symmetry in conventional Hermitian systems. For a Hermitian system, $U$ is also unitary and we have $U_{\alpha}^{-1}=U_{\alpha}^{\dagger}$.

When a Floquet system has CS, we can introduce a pair of winding numbers to characterize its topological properties~\cite{AsbothSTF,ZhouarXiv2019,ZhouDKRS2018}. In this work, we focus on the dynamical characterization of 1d Floquet systems with two bands. Without loss of generality, we assume that the Floquet operator of the system under the periodic boundary condition has the from
\begin{equation}
U(k)=e^{-ih_{y}(k)\sigma_{y}}e^{-ih_{x}(k)\sigma_{x}}.\label{eq:Uk}
\end{equation}
Here $k\in[-\pi,\pi)$ is the quasimomentum and $\sigma_{x,y}$ are Pauli matrices. $h_{x,y}(k)$ are functions of quasimomentum $k$. For a lattice model they could describe the effects of hopping and lattice potential.
We have also set the Planck constant $\hbar=1$ and driving period $T=2$.
Practically, such a Floquet evolution can be generated by a Hamiltonian
$H(k,t)$ under piecewise periodic quenches, so that
\begin{equation}
H(k,t)=\begin{cases}
h_{x}(k)\sigma_{x} & 2\ell\leq t<2\ell+1\\
h_{y}(k)\sigma_{y} & 2\ell+1<t\leq2\ell+2
\end{cases},\label{eq:Hkt}
\end{equation}
with $\ell\in\mathbb{Z}$. 
The Floquet operators in symmetric time frames can then be obtained by applying the similarity transformations $e^{-ih_x(k)\sigma_x/2}$ and $e^{-ih_y(k)\sigma_y/2}$ to $U(k)$ in Eq.~(\ref{eq:Uk}), yielding
\begin{alignat}{1}
U_{1}(k)= & e^{-i\frac{h_{x}(k)}{2}\sigma_{x}}e^{-ih_{y}(k)\sigma_{y}}e^{-i\frac{h_{x}(k)}{2}\sigma_{x}}=e^{-iH_{1}(k)},\label{eq:U1k}\\
U_{2}(k)= & e^{-i\frac{h_{y}(k)}{2}\sigma_{y}}e^{-ih_{x}(k)\sigma_{x}}e^{-i\frac{h_{y}(k)}{2}\sigma_{y}}=e^{-iH_{2}(k)}.\label{eq:U2k}
\end{alignat}
Here the effective Hamiltonians $H_{1,2}(k)$ are obtained by expanding each exponential terms in Eqs.~(\ref{eq:U1k}) and (\ref{eq:U2k}) with the Euler formula, and then recombine the relevant terms. Formally these effective Hamiltonians can be expressed as
\begin{equation}
H_{\alpha}(k)=h_{\alpha x}(k)\sigma_{x}+h_{\alpha y}(k)\sigma_{y},\label{eq:Heff}
\end{equation}
where $\alpha=1,2$ are the indices of two symmetric time frames. $h_{\alpha x}(k)$ and $h_{\alpha y}(k)$ are components of the Floquet effective Hamiltonian $H_{\alpha}(k)$~(see Appendix \ref{sec:App-00} for their explicit expressions). Their ratio defines a winding angle
\begin{equation}
\phi_{\alpha}(k)=\arctan\left[h_{\alpha y}(k)/h_{\alpha x}(k)\right].\label{eq:StatWindAngle}
\end{equation}
It is clear that the Floquet operators $U_{1,2}(k)$ in the two symmetric
time frames have the CS: $\Gamma=\sigma_{z}$, in the
sense that $\Gamma U_{1,2}(k)\Gamma=U_{1,2}^{-1}(k)$. Therefore,
following the topological characterization of chiral symmetric Floquet
systems~\cite{AsbothSTF,ZhouPRB2018}, one can introduce a pair of topological winding numbers
\begin{equation}
\nu_{\alpha}=\int_{-\pi}^{\pi}\frac{dk}{2\pi}\partial_{k}\phi_{\alpha}(k)\label{eq:W12}
\end{equation}
for $\alpha=1,2$ in the two symmetric time frames. Note that for non-Hermitian systems, the winding angle $\phi_{\alpha}(k)$ as defined in Eq.~(\ref{eq:StatWindAngle}) could be a complex number. However, its imaginary part ${\rm Im}[\phi_{\alpha}(k)]$ has no winding in the BZ $k\in[-\pi,\pi)$. So only the real part of winding angle ${\rm Re}[\phi_{\alpha}(k)]$ has a contribution to the value of $\nu_{\alpha}$ through the integral.

By recombining the winding numbers $(\nu_1,\nu_2)$, we obtain another pair of winding numbers $(\nu_{0},\nu_{\pi})$, which could fully characterize the bulk Floquet topological phases of $U(k)$~\cite{AsbothSTF,ZhouPRB2018}. Explicitly, these winding numbers are given by
\begin{equation}
\nu_{0}=\frac{\nu_{1}+\nu_{2}}{2},\qquad\nu_{\pi}=\frac{\nu_{1}-\nu_{2}}{2}.\label{eq:W0P}
\end{equation}

For Hermitian systems, the quasienergy is a phase factor with a period $2\pi/T$. When the system possesses chiral symmetry, there could be two band gaps at quasienergies $E=0$ and $E=\pi/T$. Under open boundary conditions, topologically protected edge states could appear in each gap. The integer quantized topological invariants $\nu_{0}$ and $\nu_{\pi}$ defined in Eq.~(\ref{eq:W0P}) count exactly the number of degenerate edge states at the center and boundary of the quasienergy BZ, respectively. Besides edge states, the invariants $\nu_{0}$ and $\nu_{\pi}$ are also related to different spin textures generated in the nonequilibrium evolution of the system, as will be demonstrated in Sec.~\ref{sec:QPL} of this manuscript.

For non-Hermitian systems, $\nu_{0}$ and $\nu_{\pi}$ as defined in Eq.~(\ref{eq:W0P}) may take half integer values when the conventional bulk-edge correspondence breaks down~\cite{LeePRL2016}. In this work, we focus on the dynamical characterization of bulk topological phases in non-Hermitian Floquet systems. Therefore, we ignore possible topological phase transitions due to boundary effects. Note in passing that as the two symmetric time frames are related by a $k$-dependent similarity transformation, we will have $\nu_\pi=0$ if the transformation from on frame to another does not change the topological winding pattern of $[h_{\alpha x}(k),h_{\alpha y}(k)]$ in $k$-space.

\section{Stroboscopic dynamics in biorthogonal basis}\label{sec:BioBasis}

In this section, we briefly recap the description of Floquet stroboscopic
evolution in biorthogonal representations, which will be the basis
for us to introduce our dynamical characterization. In a given time
frame $\alpha$ ($=1,2$), the right and left Floquet eigenvectors
satisfy the eigenvalue equations of the effective Hamiltonian $H_{\alpha}$ and its Hermitian conjugate, i.e.,
\begin{equation}
H_{\alpha}(k)|\psi_{s}^{\alpha}(k)\rangle=E_{s}(k)|\psi_{s}^{\alpha}(k)\rangle,
\end{equation}
and
\begin{equation}
H_{\alpha}^{\dagger}(k)|\widetilde{\psi}_{s}^{\alpha}(k)\rangle=E_{s}^{*}(k)|\widetilde{\psi}_{s}^{\alpha}(k)\rangle.
\end{equation}
Here $s=+,-$ are the indices of two Floquet bands, with quasienergy dispersions $E_{\pm}(k)\equiv\pm E(k)$. Furthermore, the left and right Floquet eigenvectors satisfy the biorthogonal normalization and completeness relations
\begin{equation}
\langle\widetilde{\psi}_{s}^{\alpha}(k)|\psi_{s'}^{\alpha}(k)\rangle=\delta_{ss'},\quad\sum_{s=\pm}|\psi_{s}^{\alpha}(k)\rangle\langle\widetilde{\psi}_{s}^{\alpha}(k)|=1.\label{eq:BiNormCond}
\end{equation}
Then according to the biorthogonal quantum mechanics~\cite{BioQM}, the stroboscopic evolutions of an arbitrary initial state in the representations of left and right eigenbasis take the forms
\begin{alignat}{1}
\langle\widetilde{\psi}^{\alpha}(k,n)|= & \sum_{s=\pm}c_{s}^{*}(k)e^{+iE_{s}^{*}(k)n}\langle\widetilde{\psi}_{s}^{\alpha}(k)|,\label{eq:PsiLn}\\
|\psi^{\alpha}(k,n)\rangle= & \sum_{s=\pm}c_{s}(k)e^{-iE_{s}(k)n}|\psi_{s}^{\alpha}(k)\rangle,\label{eq:PsiRn}
\end{alignat}
where $n\in\mathbb{N}$ counts the number of evolution periods, and the initial amplitude $c_{s}(k)=\langle\widetilde{\psi}_{s}^{\alpha}(k)|\psi^{\alpha}(k,0)\rangle=\langle\widetilde{\psi}^{\alpha}(k,0)|\psi_{s}^{\alpha}(k)\rangle$.
Note that when $E_{s}(k)\neq E_{s}^{*}(k)$, the biorthogonal normalization
condition in Eq.~(\ref{eq:BiNormCond}) may not be satisfied during the evolution.
We then need to introduce a normalization factor when evaluating the
expectation value of an operator $O$ after the evolution over $n$
driving periods in a given time frame $\alpha$, i.e.,
\begin{equation}
\langle O(k,n)\rangle_{\alpha}\equiv\frac{\langle\widetilde{\psi}^{\alpha}(k,n)|O|\psi^{\alpha}(k,n)\rangle}{\langle\widetilde{\psi}^{\alpha}(k,n)|\psi^{\alpha}(k,n)\rangle}.\label{eq:BioAve}
\end{equation}
For the effective Hamiltonians in Eq.~(\ref{eq:Heff}), the left and
right Floquet eigenvectors are explicitly given by
\begin{alignat}{1}
\langle\widetilde{\psi}_{s}^{\alpha}(k)|= & \frac{1}{\sqrt{2}E_{s}(k)}\left[h_{\alpha x}(k)+ih_{\alpha y}(k),E_{s}(k)\right],\label{eq:LeftEigVec}\\
|\psi_{s}^{\alpha}(k)\rangle= & \frac{1}{\sqrt{2}E_{s}(k)}\left[h_{\alpha x}(k)-ih_{\alpha y}(k),E_{s}(k)\right]^{\top},\label{eq:RightEigVec}
\end{alignat}
where $E_{s}(k)=s\sqrt{h_{\alpha x}^{2}(k)+h_{\alpha y}^{2}(k)}$, with
$s=+,-$, $\alpha=1,2$ and $\top$ executes matrix transpose. It
is straightforward to check that Eqs. (\ref{eq:LeftEigVec}) and (\ref{eq:RightEigVec})
satisfy the relations in Eq.~(\ref{eq:BiNormCond}).

\section{Time-averaged spin textures and dynamical winding numbers}\label{sec:AveSpinTex}

With the help of the biorthogonal formalism introduced in Sec.~\ref{sec:BioBasis}, we can now discuss how to extract the dynamical winding numbers from the stroboscopic averaged spin textures of the system. These dynamical winding numbers are further shown to be equal to the topological invariants given by Eq.~(\ref{eq:W12}) in the corresponding symmetric time frames. Their combinations then result in a dynamical characterization of Floquet topological phases in the system.

In long-time limit, the stroboscopic averaged spin textures are obtained from the biorthogonal expectation values of Pauli spin operators $\sigma_{x}$ and $\sigma_{y}$ as
\begin{equation}
r_{j}^{\alpha}(k)\equiv\lim_{N\rightarrow\infty}\frac{1}{N}\sum_{n=1}^{N}\langle\sigma_{j}(k,n)\rangle_{\alpha},\label{eq:AveSpinTex}
\end{equation}
where $j=x,y$, $\alpha=1,2$ are the indices of symmetric time frames and $N$ is the total number of driving periods.
The explicit definition of $\langle\sigma_{j}(k,n)\rangle_{\alpha}$ is given
by Eq.~(\ref{eq:BioAve}). The winding angles of averaged spin textures $[r_{x}^{\alpha}(k),r_{y}^{\alpha}(k)]$ at a given quasimomentum $k$ is defined as
\begin{equation}
\theta_{yx}^{\alpha}(k)\equiv\arctan[r_{y}^{\alpha}(k)/r_{x}^{\alpha}(k)].\label{eq:DynWindAngle}
\end{equation}
Then it can be shown that the winding angle of long-time averaged stroboscopic spin textures around the first BZ $k\in[-\pi,\pi)$ yields a dynamical winding number
\begin{equation}
W_{\alpha}\equiv\int_{-\pi}^{\pi}\frac{dk}{2\pi}\partial_{k}\theta_{yx}^{\alpha}(k),
\label{eq:WAlp}
\end{equation}
which is equal to the topological winding number $\nu_{\alpha}$ of the Floquet operator in the symmetric time frame $\alpha$ ($=1,2$) as defined by Eq.~(\ref{eq:W12}). The bulk topological invariants $(\nu_{0},\nu_{\pi})$ of the Floquet system, given by Eq.~(\ref{eq:W0P}), are then related to the dynamical winding numbers by
\begin{equation}
\nu_{0}=\frac{W_{1}+W_{2}}{2}\qquad\nu_{\pi}=\frac{W_{1}-W_{2}}{2},\label{eq:DymvsStatWindNum}
\end{equation}
which is the main result of this work.

To reach Eqs.~(\ref{eq:WAlp}) and (\ref{eq:DymvsStatWindNum}), we first plug Eqs.~(\ref{eq:PsiLn})-(\ref{eq:BioAve}) into Eq.~(\ref{eq:AveSpinTex}), which yields
\begin{equation}
r_{j}^{\alpha}=\lim_{N\rightarrow\infty}\frac{1}{N}\sum_{n=1}^{N}\frac{\sum_{ss'}D_{ss'}e^{-i\Delta_{ss'}n}\langle\widetilde{\psi}_{s'}^{\alpha}|\sigma_{j}|\psi_{s}^{\alpha}\rangle}{\sum_{s}D_{ss}e^{-i\Delta_{ss}n}},\label{eq:AveSpinTex2}
\end{equation}
where $s,s'=+,-$, $\alpha=1,2$, $j=x,y$. $N$ is the total number of driving periods and the $k$-dependence in all states and functions have been suppressed for symbolic convenience. We have also introduced compact notations for the overlapping amplitudes and spectral gaps as
\begin{alignat}{1}
D_{ss'}(k)\equiv & \,c_{s}(k)c_{s'}^{*}(k),\\
\Delta_{ss'}(k)\equiv & \,E_{s}(k)-E_{s'}^{*}(k),
\end{alignat}
where the initial state amplitude $c_s(k)$ and quasienergy dispersion $E_{s}(k)$ are defined in Sec.~\ref{sec:BioBasis}.

In the long-time limit $N\rightarrow\infty$, the dynamical winding
angle $\theta_{yx}^{\alpha}(k)$ in Eq.~(\ref{eq:DynWindAngle}) converges
to the static winding angle $\phi_{\alpha}(k)$ in Eq.~(\ref{eq:StatWindAngle})
under generic initial conditions. To see this, we first notice that if $E_{s}(k)\in\mathbb{R}$ at the quasimomentum $k$, we have
$\Delta_{ss}(k)=0$ and $\sum_{s=\pm}D_{ss}e^{-i\Delta_{ss}n}=1$ in Eq.~(\ref{eq:AveSpinTex2}). In this case, the numerator of Eq.~(\ref{eq:AveSpinTex2}) contains
static terms (with $s=s'$) and oscillating terms (with $s\neq s'$). 
The later will vanish under the summation and average over many driving periods $N$. In the meantime, using Eqs.~(\ref{eq:LeftEigVec}) and (\ref{eq:RightEigVec}), the biorthogonal expectation values of Pauli spin operators are given by
\begin{equation}
\langle\widetilde{\psi}_{s}^{\alpha}(k)|\sigma_{j}|\psi_{s}^{\alpha}(k)\rangle=h_{\alpha j}(k)/E_{s}(k),\label{eq:SpinAve}
\end{equation}
where $\alpha=1,2$, $s=+,-$ and $j=x,y$. Therefore, when the quasienergy $E_{s}(k)$
is real at quasimomentum $k$, Eq.~(\ref{eq:AveSpinTex2}) will
reduce to
\begin{equation}
r_{j}^{\alpha}(k)=\left[|c_{+}(k)|^{2}-|c_{-}(k)|^{2}\right]h_{\alpha j}(k)/E(k)\label{eq:rAlpj1}.
\end{equation}
If the initial state is prepared in a way such that $|c_{+}(k)|\neq|c_{-}(k)|$, we will have $\theta_{yx}^{\alpha}(k)=\arctan[r_{y}^{\alpha}(k)/r_{x}^{\alpha}(k)]=\arctan[h_{\alpha y}(k)/h_{\alpha x}(k)]$,
i.e., $\theta_{yx}^{\alpha}(k)=\phi_{\alpha}(k)$ according to Eqs.~(\ref{eq:rAlpj1}) and (\ref{eq:StatWindAngle}). Measuring the winding angle $\theta_{yx}^{\alpha}(k)$ of averaged spin textures then allows one to extract the static winding angle $\phi_{\alpha}(k)$ and the topological winding number $\nu_{\alpha}$ in Eq.~(\ref{eq:W12}), yielding the relations between static and dynamical winding numbers in Eq.~(\ref{eq:DymvsStatWindNum}).

The same conclusion can be drawn if the quasienergy $E_{s}(k)$ is complex at the quasimomentum $k$, which is the more typical situation in non-Hermitian systems. In this case, $r_{j}^{\alpha}(k)$ in Eq.~(\ref{eq:AveSpinTex2}) can be expressed explicitly (see Appendix \ref{sec:App-0} for the expression). Then it is clear that if ${\rm Im}(E)>0$~(${\rm Im}(E)<0$) and $|c_+(k)|\geq |c_-(k)|$~($|c_+(k)|\leq |c_-(k)|$), we obtain $r_{j}^{\alpha}(k)=\langle\widetilde{\psi}_{+}^{\alpha}(k)|\sigma_{j}|\psi_{+}^{\alpha}(k)\rangle$ ($r_{j}^{\alpha}(k)=\langle\widetilde{\psi}_{-}^{\alpha}(k)|\sigma_{j}|\psi_{-}^{\alpha}(k)\rangle$) from Eq.~(\ref{eq:AveSpinTex3}) after dropping all exponentially decaying terms in the limit $N\rightarrow\infty$.
Therefore, according to Eqs.~(\ref{eq:SpinAve}) and (\ref{eq:StatWindAngle}) we have $r_{y}^{\alpha}(k)/r_{x}^{\alpha}(k)=h_{\alpha y}(k)/h_{\alpha x}(k)$ and $\theta_{yx}^{\alpha}(k)=\phi_{\alpha}(k)$, which again allow us to extract the static winding angle $\phi_\alpha(k)$ and topological winding number $\nu_{\alpha}$ from the time-averaged spin textures even when $E_{s}(k)\notin\mathbb{R}$. Practically, due to the fast decay of irrelevant terms caused by non-vanishing imaginary parts of the quasienergies, we only need the initial conditions to be $c_{\pm}(k)\neq0$ in order to reach this conclusion.

These analysis complete our dynamical characterization of chiral symmetric non-Hermitian Floquet topological phases in 1d. Note that the relations in Eq.~(\ref{eq:DymvsStatWindNum}) only rely on simple constraints on the initial state of the dynamics, i.e., imbalanced and non-vanishing populations on both Floquet bands at each quasimomentum $k$. This reduces the demands on initial state preparations and makes it more flexible for the experimental detections of these dynamical winding numbers. In an earlier work, the relation between dynamical
and static winding numbers has also been proved in evolutions driven
by time-independent Hamiltonians. Our finding can thus be viewed as
an extension of the results in Ref.~\cite{ZhuarXiv2019} to periodically driven
systems with unique non-Hermitian Floquet topological phases.

In the following section, we introduce a non-Hermitian piecewise quenched lattice (PQL) model, in which our theory is demonstrated by direct numerical simulations.

\section{Non-Hermitian Floquet topological phases in a PQL}\label{sec:QPL}

\begin{figure}
	\begin{centering}
		\includegraphics[scale=0.5]{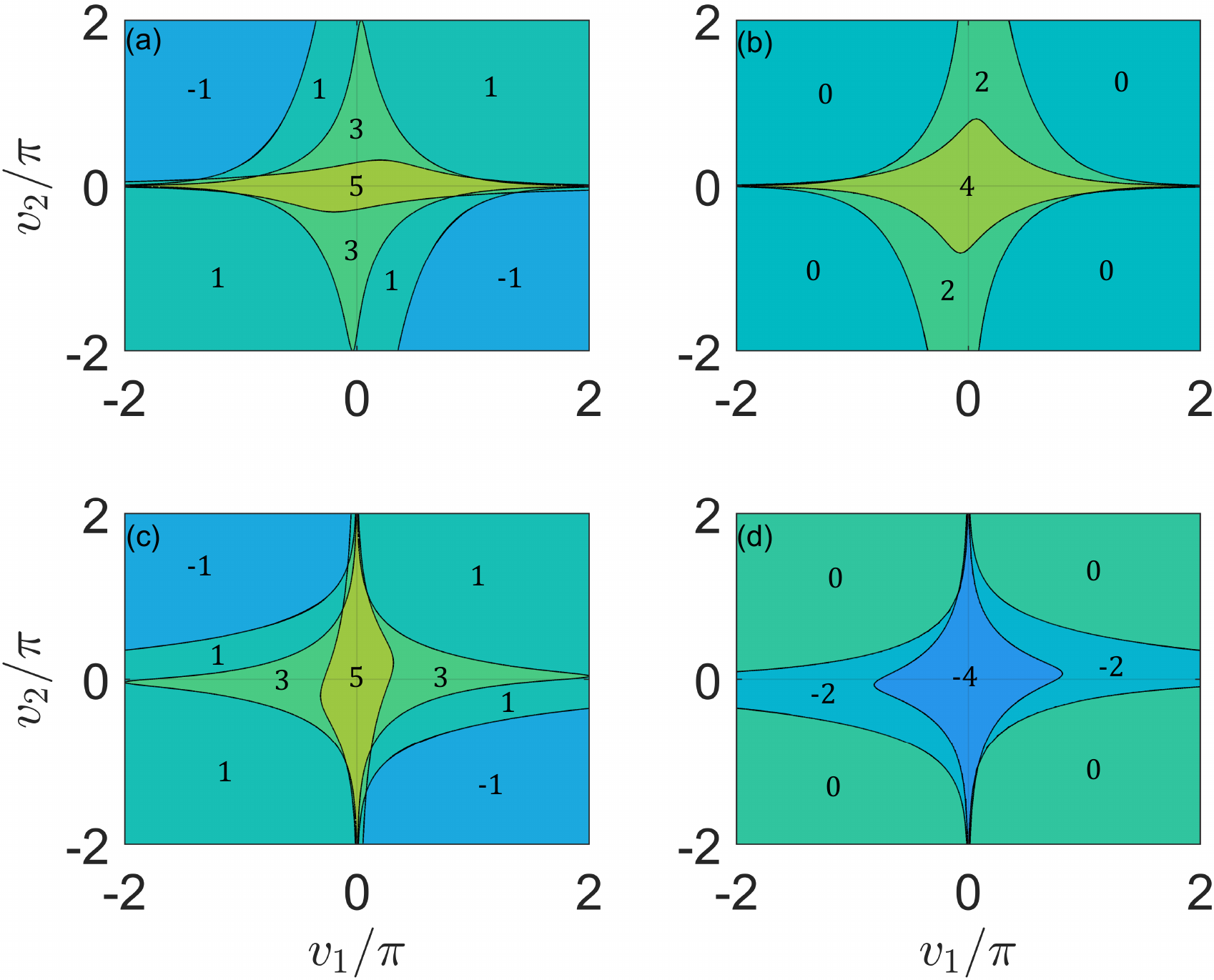}
		\par\end{centering}
	\caption{Floquet topological phase diagrams of the periodically quenched lattice model, as defined in Eq.~(\ref{eq:PQL}, versus the imaginary part of hopping amplitudes $(v_1,v_2)$. The real part of hopping amplitudes are set as $(u_1,u_2)=(0.5\pi,4.5\pi)$ for panels (a), (b) and $(u_1,u_2)=(4.5\pi,0.5\pi)$ for panels (c), (d). The values of topological winding numbers $\nu_0$ and $\nu_\pi$ are denoted in panels (a), (c) and (b), (d), respectively.	\label{fig:W_vs_v1_v2}}
\end{figure}

To demonstrate our theory, we consider a PQL model in the form of Eq.~(\ref{eq:Hkt}). The Hamiltonians in the two halves of a driving period are explicitly given by
\begin{equation}
h_{x}(k)=J_{1}\cos k\qquad h_{y}(k)=J_{2}\sin k,\label{eq:hxyk}
\end{equation}
where $k\in[-\pi,\pi)$ is the quasimomentum. $J_{1,2}=u_{1,2}+iv_{1,2}$ are hopping amplitudes between nearest neighbor unit cells of the lattice, with the real and imaginary parts being $(u_1,u_2)$ and $(v_1,v_2)$, respectively. The corresponding Floquet operator of the system is given by
\begin{equation}
U(k)=e^{-iJ_{2}\sin k\sigma_{y}}e^{-iJ_{1}\cos k\sigma_{x}}.\label{eq:PQL}
\end{equation}
In the two symmetric time frames, it admits the form $U_1(k)=e^{-iJ_{1}\cos k\sigma_{x}/2}e^{-iJ_{2}\sin k\sigma_{y}}e^{-iJ_{1}\cos k\sigma_{x}/2}$ and $U_2(k)=e^{-iJ_{2}\sin k\sigma_{y}/2}e^{-iJ_{1}\cos k\sigma_{x}}e^{-iJ_{2}\sin k\sigma_{y}/2}$. In the Hermitian limit $v_{1}=v_{2}=0$, this system has been shown
to possess rich Floquet topological phases in the context of a spin-$1/2$
double kicked rotor~\cite{ZhouDKRS2018}. With nonvanishing imaginary hopping
amplitudes $v_{1,2}\neq 0$, a series of topological phase transitions could
appear in the system. Accompanying each transition, a non-Hermitian Floquet
topological phase emerges, which can be characterized by the topological
winding numbers $(\nu_{0},\nu_{\pi})$ as defined in Eq.~(\ref{eq:W0P}).
In in Fig.~\ref{fig:W_vs_v1_v2}, we show two representative configurations of the resulting Floquet topological phase diagrams. Each region with a uniform color
in Fig.~\ref{fig:W_vs_v1_v2}(a,c) {[}(b,d){]} corresponds to a non-Hermitian Floquet
topological phase with a quantized topological invariant $\nu_{0}$
($\nu_{\pi}$), whose value is denoted in the corresponding region. The black lines denote boundaries between different topological phases.
Notably, we find non-Hermitian Floquet topological phases with large winding numbers in these phase diagrams. Their physical origin can be traced back to the long-range hopping terms in the Floquet effective Hamiltonians. According to the Baker-Campbell-Hausdorff formula, these long-range hopping terms are generated by the nested commutators between $h_x(k)$ and $h_y(k)$ in Eq~.(\ref{eq:hxyk}), which themselves only possess short-range hoppings. The realization of longe-range hopping Hamiltonians from short range ones through periodic drivings is one of the key paragidms in Floquet engineering. Our findings here demontrate that it is also applicable to non-Hermitian systems. Under open boundary conditions, the topological invariants $(\nu_0,\nu_\pi)$ predict the number of degenerate edge modes at quasienergies $0$ and $\pi$. The degeneracy of these non-Hermitian Floquet edge modes are protected by the chiral symmetry of the system~\cite{ZhouPRB2018}, which is given by $\Gamma=\sigma_z$.

\begin{figure}
	\begin{centering}
		\includegraphics[scale=0.5]{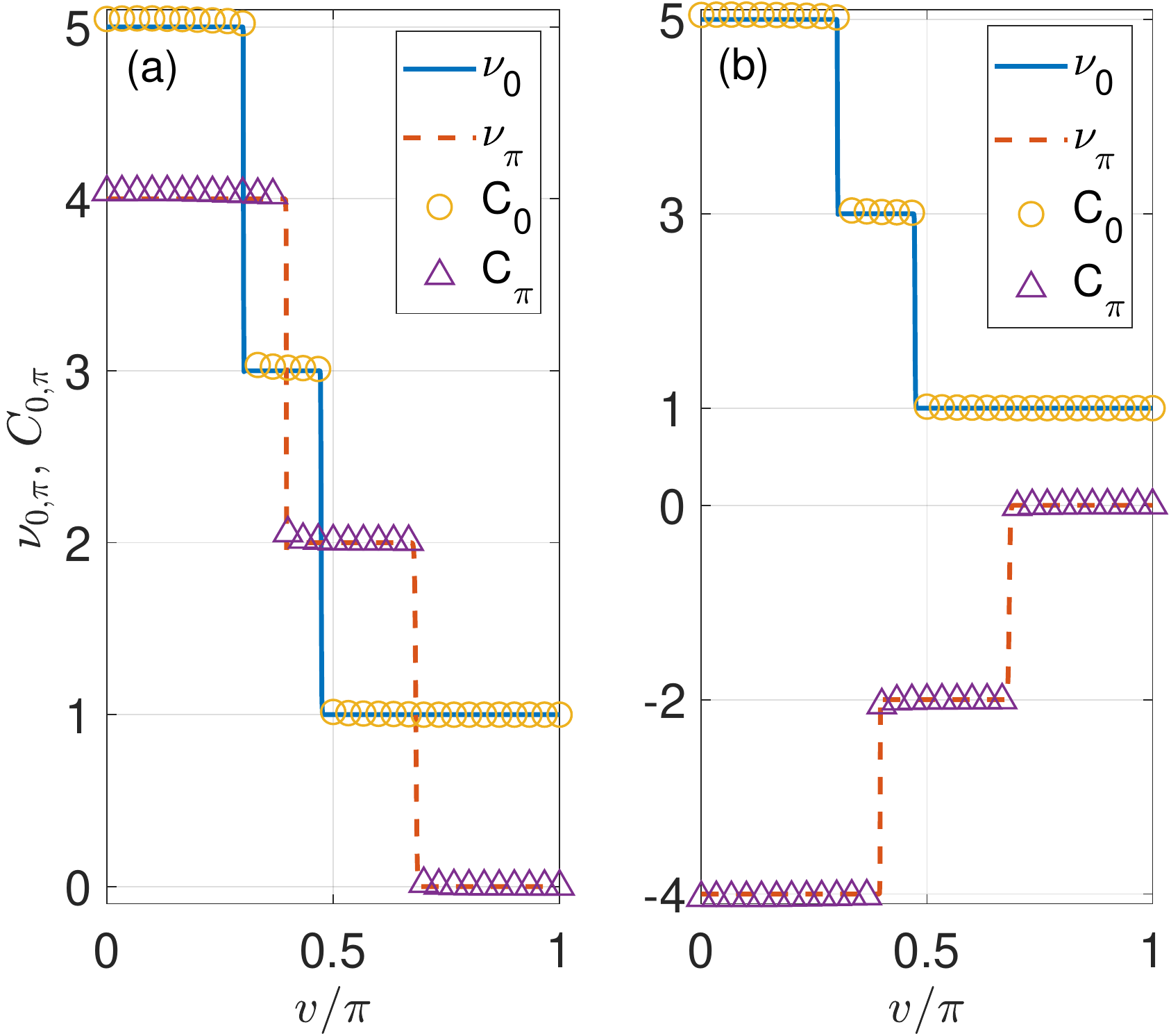}
		\par\end{centering}
	\caption{Winding numbers $\nu_{0}$ (blue solid line), $\nu_{\pi}$ (red dashed line) and mean chiral displacements $C_{0}$ (yellow circles), $C_{\pi}$ (purple triangles) versus the imaginary part of hopping amplitudes $J_{1}=u_{1}+iv$ and $J_{2}=u_{2}+iv$. System parameters are $(u_{1},u_{2})=(0.5\pi,4.5\pi)$ for panel (a) and $(u_{1},u_{2})=(4.5\pi,0.5\pi)$ for panel (b). Results for $C_{0},C_{\pi}$ are averaged over $50$ driving periods of the evolution governed by Eq.~(\ref{eq:PQL}). \label{fig:W-MCD_vs_v}}
\end{figure}

The non-Hermitian Floquet topological phases in Fig.~\ref{fig:W_vs_v1_v2} can be characterized dynamically by the mean chiral displacement~\cite{ZhouarXiv2019,ZhouDKRS2018,MCD1,MCD2,MCD3}
or the dynamical winding numbers introduced in Sec.~\ref{sec:AveSpinTex}.
In Fig.~\ref{fig:W-MCD_vs_v}, we present two examples of the winding numbers $(\nu_{0},\nu_{\pi})$
versus the imaginary parts of hopping amplitudes $v_{1}=v_{2}=v$ in
our PQL model (\ref{eq:PQL}). In both cases,
we observe that the topological invariants $\nu_{0}$ (solid line)
and $\nu_{\pi}$ (dashed line) and their changes are consistent with the mean chiral
displacements $C_{0}$ (circles) and $C_{\pi}$ (triangles), respectively (see Appendix \ref{sec:App-A} for the definitions of $C_{0,\pi}$).

\begin{figure}
	\begin{centering}
		\includegraphics[scale=0.5]{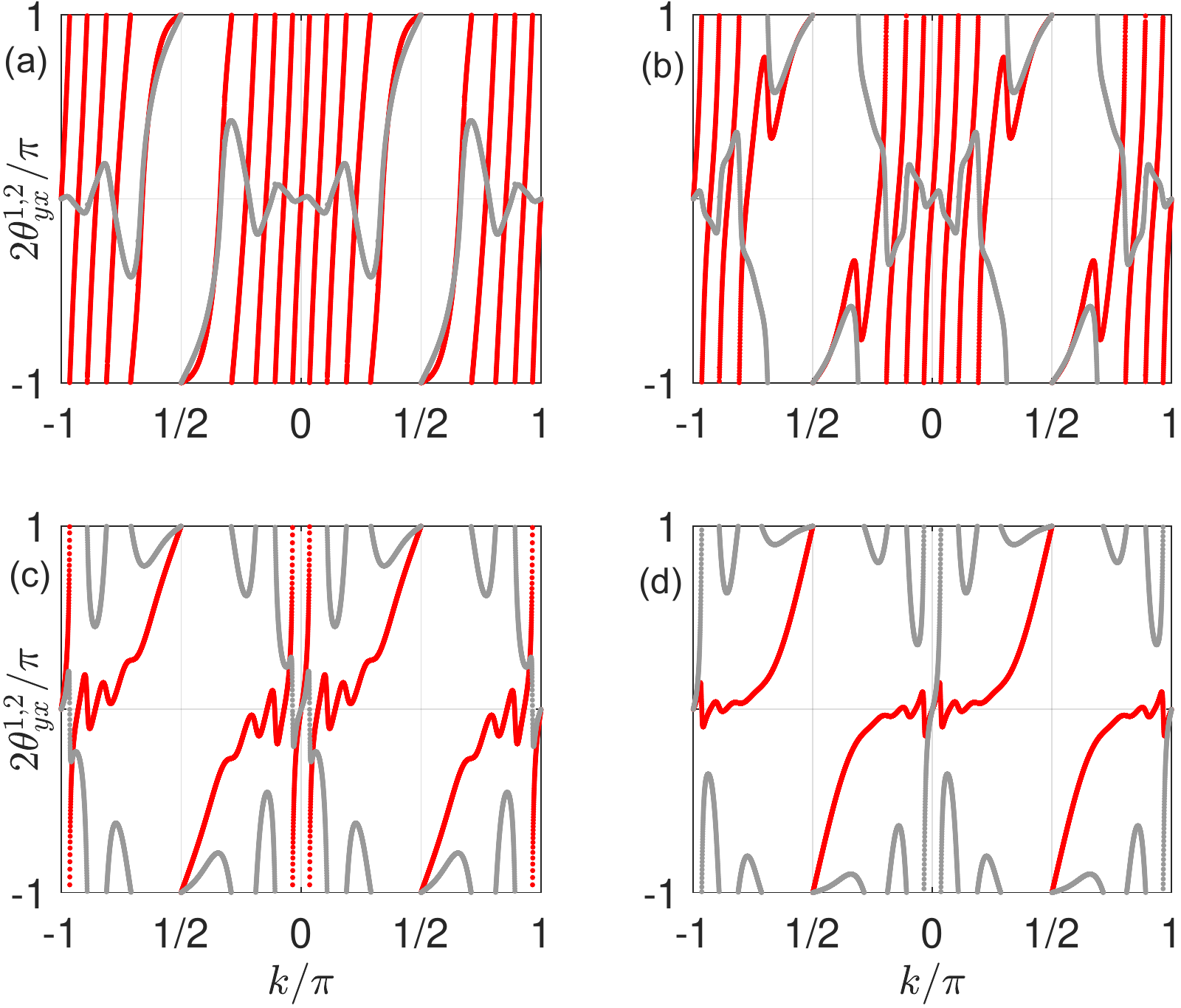}
		\par\end{centering}
	\caption{Winding angles $\theta_{yx}^1$ (red dots) and $\theta_{yx}^2$ (gray dots) of the time-averaged spin textures versus the quasimomentum $k$. The real parts of hopping amplitudes are $(u_1,u_2)=(0.5\pi,4.5\pi)$ for all panels. The imaginary parts of hopping amplitudes are $v_1=v_2=v=0.2\pi,0.35\pi,0.6\pi,0.9\pi$ for panels (a)-(d). The dynamical winding numbers, derived from the winding angles around the first BZ are $(W_1,W_2)=(9,1), (7,-1), (3,-1), (1,1)$, yielding $(\frac{W_1+W_2}{2},\frac{W_1-W_2}{2})=(5,4), (3,4), (1,2), (1,0)$ for panels (a)- (d). They are consistent with the theoretical values of $(\nu_0,\nu_\pi)$ in Fig.~\ref{fig:W-MCD_vs_v}(a) at the corresponding system parameters.}\label{fig:WA1}
\end{figure}

\begin{figure}
	\begin{centering}
		\includegraphics[scale=0.5]{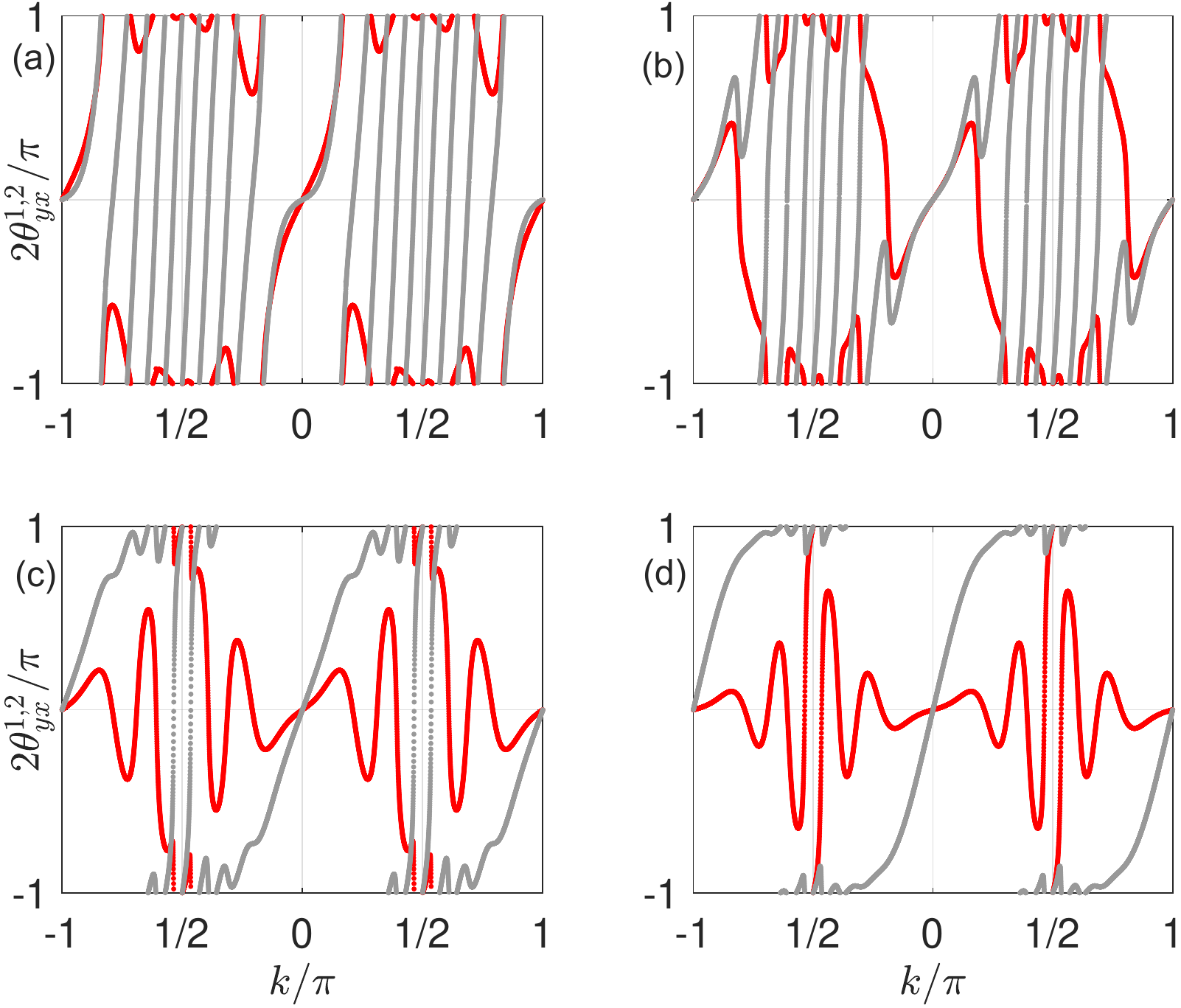}
		\par\end{centering}
	\caption{Winding angles $\theta_{yx}^1$ (red dots) and $\theta_{yx}^2$ (gray dots) of the time-averaged spin textures versus the quasimomentum $k$. The real parts of hopping amplitudes are $(u_1,u_2)=(4.5\pi,0.5\pi)$ for all panels. The imaginary parts of hopping amplitudes are $v_1=v_2=v=0.2\pi,0.35\pi,0.6\pi,0.9\pi$ for panels (a)-(d). The dynamical winding numbers, derived from the winding angles around the first BZ are $(W_1,W_2)=(1,9), (-1,7), (-1,3), (1,1)$, yielding $(\frac{W_1+W_2}{2},\frac{W_1-W_2}{2})=(5,-4), (3,-4), (1,-2), (1,0)$ for panels (a)-(d). They are consistent with the theoretical values of $(\nu_0,\nu_\pi)$ in Fig.~\ref{fig:W-MCD_vs_v}(b) at the corresponding system parameters.}\label{fig:WA2}
\end{figure}

In Figs.~\ref{fig:WA1} and \ref{fig:WA2}, we further compute the dynamical winding angles
$\theta_{yx}^{1,2}(k)$ as defined in Eq.~(\ref{eq:DynWindAngle})
for several different Floquet topological phases. 
The panels in Fig.~\ref{fig:WA1} are obtained at four sets of system parameters in Fig.~\ref{fig:W-MCD_vs_v}(a), corresponding to four distinct non-Hermitian Floquet topological phases.
We see that the winding angles of time-averaged spin textures $\theta_{yx}^{1,2}(k)$ around the first BZ yield the dynamical winding numbers $(W_{1},W_{2})=(9,1)$,
$(7,-1)$, $(3,-1)$ and $(1,1)$ at four different sets of system parameters, according to Eq.~(\ref{eq:WAlp}).
Their combinations, in terms of Eq.~(\ref{eq:DymvsStatWindNum}),
give the correct winding numbers of the corresponding non-Hermitian
Floquet topological phases in Fig.~\ref{fig:W-MCD_vs_v}(a). Similarly, counting the winding angles of time-averaged spin textures in Fig.~\ref{fig:WA2} yields the dynamical
winding numbers $(W_{1},W_{2})=(1,9)$, $(-1,7)$, $(-1,3)$ and $(1,1)$
at other four different sets of system parameters. Their combinations also
produce the winding numbers of the corresponding non-Hermitian Floquet
topological phases in Fig.~\ref{fig:W-MCD_vs_v}(b).
In all the calculations of winding angles, the dynamical spin textures are averaged over $N=100$ driving periods. We have also verified that a good convergence between theoretical and numerical results can be achieved by an average over $N=30$ driving periods in most of the examples we investigated, which should be within reach in current experimental situations.
With these numerical evidence, we
conclude that the winding numbers of time-averaged spin textures could
indeed provide a dynamical characterization for non-Hermitian Floquet
topological phases with CS in 1d.

Some distinctions between our dynamical approach and the exisiting approaches for static systems~\cite{ZhuarXiv2019} deserve to be mentioned. First, in Floquet systems the dynamical winding numbers can be obtained from the stroboscopic average of spin textures. In contrast, in static systems the spin textures need to be averaged over a continuous time duration for the construction of dynamical winding numbers, which may pose more challenges to experiments. Second, a chiral symmetric Floquet topological phase is characterized by a pair of topological invariants. Therefore, one needs to prepare the initial states and measure the dynamical spin textures in two complementary symmetric time frames. This is more complicated then the situations in static systems, where conducting the measurements in an arbitrary time frame is sufficient. Nevertheless, we have shown that non-Hermitian Floquet systems possess rich topological phases. Our approach provides a powerful dynamical tool in decoding their topological properties.

\section{Summary}

In this work, we propose an approach to dynamically characterize non-Hermitian
Floquet topological phases in 1d with chiral symmetry. We utilize the relative
phase between two long-time averaged spin components to define a dynamical
winding number, which is coincide with the topological invariant of
the Floquet operator. The effectiveness of our approach is verified
in a piecewise periodically quenched lattice model with rich non-Hermitian
Floquet topological phases. We also compared our approach with another
dynamical characterization strategy based on the mean chiral displacement
of a wavepacket, and obtain consistent results.
In experiments, the dynamical winding numbers of spin textures studied in this work might be directly measurable in existing optical~\cite{Exp13} and photonic~\cite{Exp23} setups with dissipation and losses.

In future work, it would be interesting to consider the extension of our dynamical approaches to two dimensions for characterizing Chern and second order Floquet topological insulators in both Hermitian and non-Hermitian settings.
For multiple-band models with chiral symmetry, the spin textures are formed by expectation values of the Dirac $\gamma$ matrices~\cite{SpinTex1}, and our approach should also be applicable after slight modifications. In recent studies, it has been found that due to the difference between transpose and complex conjugate of non-Hermitian matrices, non-Hermtian topological phases form an enlarged ``periodic table'' compared with their Hermitian counterparts~\cite{Class7}. It is expected that periodic drivings would further enlarge this periodic table, and the generalizations of our dynamical approach to differet symmetry classes in this table deserve detailed investigations. 
Furthermore, exploring the robustness of our approach to disorder, nonlinear or many-body interactions and other kinds of environmental effects should also be important for its applications in different types of non-Hermitian Floquet systems.


\section*{Acknowledgement}
This work is supported by the the National Natural Science Foundation of China (Grant No. 11905211), the China Postdoctoral Science Foundation (Grant No. 2019M662444), the Fundamental Research Funds for the Central Universities (Grant No. 841912009), the Young Talents Project at Ocean University of China (Grant No. 861801013196), and the Applied Research Project of Postdoctoral Fellows in Qingdao (Grant No. 861905040009).

\appendix
\section{Explicit expressions of the components of Floquet effective Hamiltonians}\label{sec:App-00}
By expanding $U_1(k),U_2(k)$ in the main text and recombining the relevant terms, we find
\begin{alignat}{1}
h_{1x}(k)= & E(k)\sin[h_{x}(k)]\cos[h_{y}(k)]/\sin[E(k)],\\
h_{1y}(k)= & E(k)\sin[h_{y}(k)]/\sin[E(k)],\\
h_{2x}(k)= & E(k)\sin[h_{x}(k)]/\sin[E(k)],\\
h_{2y}(k)= & E(k)\cos[h_{x}(k)]\sin[h_{y}(k)]/\sin[E(k)],
\end{alignat}
where $E(k)=\arccos(\cos[h_x(k)]\cos[h_y(k)])$ is the quasienergy dispersion.

\section{Explicit expression of $r_{j}^{\alpha}(k)$}\label{sec:App-0}
When the quasienergy $E_s(k)$ is complex, Eq.~(\ref{eq:AveSpinTex2}) in the main text can be expressed explicitly as
\begin{alignat}{1}
r_{j}^{\alpha}(k)= & \lim_{N\rightarrow\infty}\frac{1}{N}\sum_{n=1}^{N}\left[\frac{D_{++}e^{2\eta n}\langle\sigma_{j}\rangle_{++}^{\alpha}+D_{--}e^{-2\eta n}\langle\sigma_{j}\rangle_{--}^{\alpha}}{D_{++}e^{2\eta n}+D_{--}e^{-2\eta n}}\right.\nonumber \\
& \left.+\frac{D_{+-}e^{-i2\lambda n}\langle\sigma_{j}\rangle_{-+}^{\alpha}+D_{-+}e^{i2\lambda n}\langle\sigma_{j}\rangle_{+-}^{\alpha}}{D_{++}e^{2\eta n}+D_{--}e^{-2\eta n}}\right]
\label{eq:AveSpinTex3}
\end{alignat}
where $\lambda\equiv{\rm Re}(E_{+})$, $\eta\equiv{\rm Im}(E_{+})$, and $\langle\sigma_{j}\rangle_{ss'}^{\alpha}=\langle\widetilde{\psi}_{s'}^{\alpha}|\sigma_{j}|\psi_{s}^{\alpha}\rangle$ for $s,s'=+,-$, $j=x,y$ and $\alpha=1,2$.

\section{Definition of the mean chiral displacement}\label{sec:App-A}
The mean chiral displacements $(C_0,C_\pi)$ in Fig.~\ref{fig:W-MCD_vs_v} of the main text are given by
\begin{equation}
C_0=\frac{C_1+C_2}{2},\qquad C_\pi=\frac{C_1-C_2}{2},\label{C0P}
\end{equation}
where the mean chiral displacements $C_{1,2}$ in the two symmetric time frames are defined as
\begin{equation}
C_{\alpha}=\lim_{N\rightarrow\infty}\frac{1}{N}\sum_{n=1}^{N}\int_{-\pi}^{\pi}\frac{dk}{2\pi}\frac{{\rm Tr}\left[\tilde{U}_{\alpha}^{\dagger n}(k)\Gamma i\partial_{k}U_{\alpha}^{n}(k)\right]}{{\rm Tr}\left[\tilde{U}_{\alpha}^{\dagger n}(k)U_{\alpha}^{n}(k)\right]}.\label{eq:MCD}
\end{equation}
Here $\Gamma$ is the CS operator, which is $\sigma_z$ for our piecewise quenched lattice model. $N$ is the total number of driving periods. $U_\alpha(k)$ is the Floquet operator in the symmetric time frame $\alpha=1,2$, and $\tilde{U}_{\alpha}(k)=\sum_{s} e^{-iE_s(k)}|\widetilde{\psi}_{s}^{\alpha}(k)\rangle\langle\psi_{s}^{\alpha}(k)|$ governs the dynamics of left vectors in the biorthogonal basis. The trace corresponds to taking the expectation value over an intial state with equal populations on both sublattice sites in the central unit cell of the lattice. In Ref.~\cite{ZhouarXiv2019}, it is proved that for a chiral symmetric non-Hermtian Floquet system in 1d, its topological winding numbers $(\nu_0,\nu_\pi)$ are equal to $(C_0,C_\pi)$. Therefore, the mean chiral displacements $(C_1,C_2)$ provide an alternative dynamical probe to the non-Hermitian topological phases of chiral symmetric Floquet systems in 1d.



\begin{thebibliography}{99}
	
	\bibitem{NHTPReview1} Z. Gong, Y. Ashida, K. Kawabata, K. Takasan, S. Higashikawa, M. Ueda, Phys. Rev. X {\bf 8}, 031079 (2018).
	
	\bibitem{NHTPReview2} H. Shen, B. Zhen, and L. Fu, Phys. Rev. Lett. {\bf 120}, 146402 (2018).
	
	\bibitem{NHTPReview3} V. M. M. Alvarez, J. E. B. Vargas,	M. Berdakin, and L. E. F. Foa Torres, Eur. Phys. J. Special Topics {\bf 227}, 1295 (2018).
	
	\bibitem{NHTPReview4} A. Ghatak, and T. Das, J. Phys.: Condens. Matter {\bf 31}, 263001 (2019).
	
	\bibitem{NHTPReview5} L. E. F. Foa Torres, J. Phys.: Mater. {\bf 3}, 014002 (2020).
	
	\bibitem{NHTPReview6} H. Zhao and L. Feng, National Science Review {\bf 5}, 183-199 (2018).
	
	\bibitem{NHTPReview7} H. Cao and J. Wiersig, Rev. Mod. Phys. {\bf 87}, 61 (2015).
	
	\bibitem{NHTPReview8} S. Longhi, EPL {\bf 120}, 64001 (2017).
	
	\bibitem{NHTPReview9} S. K. \"Ozdemir, S. Rotter, F. Nori, and L. Yang, Nature Materials {\bf 18}, 783-798 (2019).
	
	\bibitem{NHTPReview10} M. Miri and A. Alu, Science {\bf 363}, eaar7709 (2019).
	
	\bibitem{RudnerPRL2009} M. S. Rudner, and L. S. Levitov, Phys. Rev. Lett. {\bf 102}, 065703 (2009).
	
	\bibitem{EsakiPRB2011} K. Esaki, M. Sato, K. Hasebe, and M. Kohmoto, Phys. Rev. B {\bf 84}, 205128 (2011).
	
	\bibitem{DiehlNP2011} S. Diehl, E. Rico, M. A. Baranov, and P. Zoller, Nat. Phys. {\bf 7}, 971-977 (2011).
	
	\bibitem{ZhuRPA2014} B. Zhu, R. L\"u, and S. Chen Phys. Rev. A {\bf 89} 062102 (2014).
	
	\bibitem{MalzardPRL2015} S. Malzard, C. Poli, and H. Schomerus, Phys. Rev. Lett. {\bf 115}, 200402 (2015)
	
	\bibitem{YucePLA2015} C. Yuce, Phys. Lett. A {\bf 379}, 1213 (2015).
	
	\bibitem{San-Jose2016} P. San-Jose, J. Cayao, E. Prada, and R. Aguado, Sci. Rep. {\bf 6}, 21427 (2016).
	
	\bibitem{LeePRL2016} T. E. Lee, Phys. Rev. Lett. {\bf 116}, 133903 (2016).
	
	\bibitem{LeykamPRL2017} D. Leykam, K. Y. Bliokh, C. Huang, Y. Chong, and F. Nori, Phys. Rev. Lett. {\bf 118}, 040401 (2017).
	
	\bibitem{JinPRA2017} L. Jin, Phys. Rev. A {\bf 96}, 032103 (2017).
	
	\bibitem{XuPRL2017} Y. Xu, S. Wang, and L.-M. Duan, Phys. Rev. Lett. 118, 045701 (2017).
	
	\bibitem{KlettPRA2017} M. Klett, H. Cartarius, D. Dast, J. Main, G. Wunner, Phys. Rev. A {\bf 95}, 053626 (2017).
	
	\bibitem{YinPRA2018} C. Yin, H. Jiang, L. Li, R. L?u, S. Chen, Phys. Rev. A {\bf 97}, 052115 (2018).
	
	\bibitem{LieuPRB2018} S. Lieu, Phys. Rev. B {\bf 97}, 045106 (2018).
	
	\bibitem{DangelPRA2018} F. Dangel, M. Wagner, H. Cartarius, J. Main, G. Wunner, Phys. Rev. A 98, 013628 (2018).
	
	\bibitem{ZhouPRB2018} L. Zhou and J. Gong, Phys. Rev. B {\bf 98} 205417 (2018).
	
	\bibitem{ZhouPRA2018} L. Zhou, Q. Wang, H. Wang, J. Gong, Phys. Rev. A {\bf 98}, 022129 (2018).
	
	\bibitem{LiuPRL2019} T. Liu, Y. Zhang, Q. Ai, Z. Gong, K. Kawabata, M. Ueda, and F. Nori, Phys. Rev. Lett. {\bf 122}, 076801 (2019).
	
	\bibitem{LonghiPRL2019} S. Longhi,	Phys. Rev. Lett. {\bf 122}, 237601 (2019).
	
	\bibitem{ZhangPRA2019} X. Z. Zhang and Z. Song, Phys. Rev. A {\bf 99}, 012113 (2019).
	
	\bibitem{HirsbrunnerPRB2019} M. R. Hirsbrunner, T. M. Philip, and M. J. Gilbert, Phys. Rev. B {\bf 100}, 081104 (2019).
	
	\bibitem{ZhouOP2019} H. Zhou, J. Y. Lee, S. Liu, and B. Zhen, Optica {\bf 6}, 190 (2019).
	
	\bibitem{OkugawaPRB2019} R. Okugawa and T. Yokoyama, Phys. Rev. B {\bf 99}, 041202 (2019).
	
	\bibitem{YangPRB2019} Z. Yang and J. Hu, Phys. Rev. B {\bf 99}, 081102 (2019).
	
	\bibitem{YoshidaPRB2019} T. Yoshida, R. Peters, N. Kawakami, and Y. Hatsugai, Phys. Rev. B {\bf 99}, 121101 (2019).
	
	\bibitem{MoorsPRB2019} K. Moors, A. A. Zyuzin, A. Y. Zyuzin, R. P. Tiwari, and T. L. Schmidt, Phys. Rev. B {\bf 99}, 041116 (2019).
	
	\bibitem{BudichPRB2019} J. C. Budich, J. Carlstr\"om, F. K. Kunst, and E. J. Bergholtz, Phys. Rev. B {\bf 99}, 041406 (2019).
	
	\bibitem{CaspelSP2019} M. T. van Caspel, S. E. T. Arze, and I. P. Castillo, SciPost Phys. {\bf 6}, 026 (2019).
	
	%
	
	\bibitem{NHSkin1} S. Yao, Z. Wang, Phys. Rev. Lett. {\bf 121}, 086803 (2018).
	
	\bibitem{NHSkin2} H. Jiang, L. Lang, C. Yang, S. Zhu, and S. Chen, Phys. Rev. B {\bf 100}, 054301 (2019).
	
	\bibitem{NHSkin3} V. M. Martinez Alvarez, J. E. Barrios Vargas, and L. E. F. Foa Torres, Phys. Rev. B {\bf 97}, 121401 (2018).
	
	\bibitem{NHSkin4} S. Yao, F. Song, and Z. Wang, Phys. Rev. Lett. {\bf 121}, 136802 (2018).
	
	\bibitem{NHSkin5} C. H. Lee and R. Thomale, Phys. Rev. B {\bf 99}, 201103 (2019).
	
	\bibitem{NHSkin6} K. Yokomizo and S. Murakami, Phys. Rev. Lett. {\bf 123}, 066404 (2019).
	
	\bibitem{NHSkin7} T. Deng and W. Yi, Phys. Rev. B {\bf 100}, 035102 (2019).
	
	\bibitem{NHSkin8} F. Song, S. Yao, and Z. Wang, Phys. Rev. Lett. {\bf 123}, 170401 (2019).
	
	\bibitem{NHSkin9} S. Longhi, Phys. Rev. Research {\bf 1}, 023013 (2019).
	
	\bibitem{NHSkin10} N. Okuma and M. Sato, Phys. Rev. Lett. {\bf 123}, 097701 (2019).
	
	\bibitem{NHSkin11} C.H. Lee, L. Li, and J.B. Gong, Phys. Rev. Lett. {\bf 123}, 016805 (2019).
	
	\bibitem{NHSkin12} X. Zhang and J. Gong, arXiv:1909.10234 (2019).
	
	\bibitem{NHES1} L. Herviou, N. Regnault, and J. H. Bardarson, arXiv:1908.09852 (2019).
	
	\bibitem{NHES2} P. Chang, J. You, X. Wen, and S. Ryu, arXiv:1909.01346 (2019).
	
	\bibitem{NHES3} S. Chakraborty and A. K. Sarma, arXiv:1906.00222 (2019).
	
	\bibitem{Class1} C. Liu, H. Jiang, and S. Chen, Phys. Rev. B {\bf 99}, 125103 (2019).
	
	\bibitem{Class2} K. Kawabata, S. Higashikawa, Z. Gong, Y. Ashida and M. Ueda, Nat. Commun. {\bf 10}, 297 (2019).
	
	\bibitem{Class3} Z. Ge, Y. Zhang, T. Liu, S. Li, H. Fan, and F. Nori, Phys. Rev. B {\bf 100}, 054105 (2019).
	
	\bibitem{Class4} K. Kawabata, T. Bessho, and M. Sato, Phys. Rev. Lett. {\bf 123}, 066405 (2019).
	
	\bibitem{Class5} L. Li, C. H. Lee, and J. Gong, Phys. Rev. B {\bf 100}, 075403 (2019).
	
	\bibitem{Class6} H. Zhou and J. Y. Lee, Phys. Rev. B {\bf 99} 235112 (2019).
	
	\bibitem{Class7} K. Kawabata, K. Shiozaki, M. Ueda, and M. Sato, Phys. Rev. X {\bf 9}, 041015 (2019).
	
	\bibitem{Class8} J. Y. Lee, J. Ahn, H. Zhou, and A. Vishwanath, Phys. Rev. Lett. {\bf 123}, 206404 (2019).
	
	\bibitem{Class9} W. B. Rui, Y. X. Zhao, and A. P. Schnyder, Phys. Rev. B {\bf 99}, 241110 (2019).
	
	\bibitem{Class10} C. Liu and S. Chen, Phys. Rev. B {\bf 100}, 144106 (2019).
	
	\bibitem{BBC1} F. K. Kunst, E. Edvardsson, J. C. Budich, and E. J. Bergholtz, Phys. Rev. Lett. {\bf 121}, 026808 (2018).
	
	\bibitem{BBC2} D. S. Borgnia, A. J. Kruchkov, and R. Slager, arXiv:1902.07217 (2019).
	
	\bibitem{BBC3} E. Edvardsson, F. K. Kunst, and E. J. Bergholtz, Phys. Rev. B {\bf 99}, 081302 (2019).
	
	\bibitem{BBC4} L. Jin and Z. Song, Phys. Rev. B {\bf 99}, 081103 (2019).
	
	\bibitem{BBC5} H. Wang, J. Ruan, and H. Zhang, Phys. Rev. B {\bf 99}, 075130 (2019).
	
	\bibitem{BBC6} H. Zirnstein, G. Refael, and B. Rosenow, arXiv:1901.11241 (2019).
	
	\bibitem{BBC7} L. Herviou, J. H. Bardarson, and N. Regnault, Phys. Rev. A {\bf 99}, 052118 (2019).
	
	\bibitem{BBC8} K. Imura and Y. Takane, Phys. Rev. B {\bf 100}, 165430 (2019).
	
	\bibitem{BBC9} F. K. Kunst and V. Dwivedi, Phys. Rev. B {\bf 99}, 245116 (2019).
	
	\bibitem{BBC10} F. Song, S. Yao, and Z. Wang, arXiv:1905.02211 (2019).
	
	\bibitem{BBC11} Y. Xiong, J. Phys. Commun. {\bf 2}, 035043 (2018).
	
	\bibitem{Exp11} J. M. Zeuner, M. C. Rechtsman, Y. Plotnik, Y. Lumer, S. Nolte, M. S. Rudner, M. Segev, and A. Szameit, Phys. Rev. Lett. {\bf 115}, 040402 (2015).
	
	\bibitem{Exp12} M. Hafezi, E. Demler, M. Lukin, J. Taylor, Nat. Phys. {\bf 7}, 907 (2011).
	
	\bibitem{Exp13} P. Peng, W. Cao, C. Shen, W. Qu, J. Wen, L. Jiang and Y. Xiao, Nat. Phys. {\bf 12}, 1139-1145 (2016).
	
	\bibitem{Exp21} S. Weimann, M. Kremer, Y. Plotnik, Y. Lumer, S. Nolte, K. G. Makris, M. Segev, M. C. Rechtsman, and A. Szameit, Nature Mater. {\bf 16}, 433 (2016).
	
	\bibitem{Exp22} L. Xiao, T. Deng, K. Wang, G. Zhu, Z. Wang, W. Yi, and P. Xue, arXiv:1907.12566 (2019).
	
	\bibitem{Exp23} K. Wang, X. Qiu, L. Xiao, X. Zhan, Z. Bian, B. C. Sanders, W. Yi, and P. Xue, Nature Communication {\bf 10}, 2293 (2019).
	
	\bibitem{Exp31} T. Helbig, T. Hofmann, S. Imhof, M. Abdelghany, T. Kiessling, L. W. Molenkamp, C. H. Lee, A. Szameit, M. Greiter and R. Thomale, arXiv:1907.11562 (2019).
	
	\bibitem{Exp41} C. Poli, M.	Bellec, U. Kuhl, F. Mortessagne, and H. Schomerus, Nature Commun. {\bf 6}, 6710 (2015).
	
	\bibitem{Exp42} H. Xu, D. Mason, L. Jiang, J.G.E. Harris, Nature (London) {\bf 537}, 80-83 (2016).
	
	\bibitem{Exp51} A. Ghatak, M. Brandenbourger, J. v. Wezel, and C. Coulais, arXiv:1907.11619 (2019).
	
	\bibitem{Exp52} W. Zhu, X. Fang, D. Li, Y. Sun, Y. Li, Y. Jing, and H. Chen, Phys. Rev. Lett. {\bf 121}, 124501 (2018).
	
	\bibitem{EPTransfer1} H. Xu, D. Mason, L. Jiang, and J. G. E. Harris, Nature {\bf 537}, 80-83 (2016).
	
	\bibitem{TPLaser1} G. Harari, M. A. Bandres, Y. Lumer, M. C. Rechtsman, Y. D. Chong, M. Khajavikhan, D. N. Christodoulides, and M. Segev, Science {\bf 359}, eaar4003 (2018).
	
	\bibitem{TPLaser2} M. A. Bandres, S. Wittek, G. Harari, M. Parto, J. Ren, M. Segev, D. N. Christodoulides, and M. Khajavikhan, Science {\bf 359}, eaar4005 (2018).
	
	\bibitem{TPLaser3} Y. V. Kartashov and D. V. Skryabin, Phys. Rev. Lett. {\bf 122}, 083902 (2019).
	
	\bibitem{EPSense1} Q. Zhong, J. Ren, M. Khajavikhan, D. N. Christodoulides, S. K. \"Ozdemir, and R. El-Ganainy, Phys. Rev. Lett. {\bf 122}, 153902 (2019).
	
	\bibitem{EPSense2} M. Zhang, W. Sweeney, C. W. Hsu, L. Yang, A. D. Stone, and L. Jiang, Phys. Rev. Lett. {\bf 123}, 180501 (2019).
	
	\bibitem{EPSense3} W. Chen, K. Ozdemir, G. Zhao, J. Wiersig, and L. Yang, Nature {\bf 548}, 192-196 (2017).
	
	\bibitem{EPSense4} H. Hodaei, A. U. Hassan, S. Wittek, H. Garcia-Gracia, R. El-Ganainy, D. N. Christodoulides, and M. Khajavikhan. Nature {\bf 548}, 187-191 (2017).
	
	\bibitem{ZhouarXiv2019} L. Zhou and J. Pan, Phys. Rev. A {\bf 100}, 053608 (2019).
	
	\bibitem{MCD1} F. Cardano, A. D\textquoteright Errico, A. Dauphin,
	M. Maffei, B. Piccirillo, C. de Lisio, G. D. Filippis, V. Cataudella,
	E. Santamato, L. Marrucci, M. Lewenstein and P. Massignan, Nat. Commun.
	\textbf{8}, 15516 (2017).
	
	\bibitem{MCD2} M. Maffei, A. Dauphin, F. Cardano, M. Lewenstein
	and P. Massignan, New J. Phys. \textbf{20}, 013023 (2018).
	
	\bibitem{MCD3} E. J. Meier, F. A. An, A. Dauphin, M. Maffei, P. Massignan,
	T. L. Hughes, and B. Gadway, Science {\bf 362}, 929 (2018).
	
	\bibitem{SpinTex1} L. Zhang, L. Zhang, S. Niu, and X. Liu, Science Bulletin {\bf 63}, 1385-1391 (2018).
	
	\bibitem{ZhuarXiv2019} B. Zhu, Y. Ke, H. Zhong, and C. Lee, arXiv:1907.11348 (2019).
	
	\bibitem{AsbothSTF} J. K. Asb$\acute{{\rm o}}$th, Phys. Rev. B \textbf{86},	195414 (2012); J. K. Asb$\acute{{\rm o}}$th, and H. Obuse, Phys. Rev. B \textbf{88}, 121406 (2013).
	
	\bibitem{ZhouDKRS2018} L. Zhou, J. Gong, Phys. Rev. A {\bf 97}, 063603 (2018).
	
	\bibitem{BioQM} D. C. Brody, J. Phys. A: Math. Theor. {\bf 47} 035305 (2014).
	
	
	
\end{thebibliography}
\end{document}